\documentclass{eas_Arena3}
\usepackage{graphicx}
\begin{document}

\title{Astrochemistry and Astrophotonics \\ for an Antarctic Observatory} 
\runningtitle{Astrochemistry and astrophotonics for Antarctica} 
\author{A. Kelz}
\address{Astrophysikalisches Institut Potsdam,
		An der Sternwarte 16,
		14482 Potsdam, Germany
		\email{akelz@aip.de}}
\author{M. M. Roth$^1$} 
\author{H.-G. L\"ohmannsr\"oben}
\address{Universit\"at Potsdam, 
Institut f\"ur Chemie, 
Karl-Liebknecht-Str. 25, 
14476 Potsdam-Golm} 
\author{M. Kumke$^2$}
\begin{abstract}
Due to its location and climate, Antarctica offers unique conditions
for long-period observations across a broad wavelength regime,
where important diagnostic lines for molecules and ions can be found,
that are essential to understand the chemical properties of the interstellar
medium. In addition to the natural benefits of the site, new
technologies, resulting from astrophotonics, may allow miniaturised instruments,
that are easier to winterise and advanced filters to further
reduce the background in the infrared. 
\end{abstract}
\maketitle
\section{Introduction}
An observatory in Antarctica offers the potential to study many astrochemical
signatures, such as diffuse interstellar bands (DIBs) and molecular properties of
the interstellar medium (ISM). Given, that Antarctica offers one of the best atmospheric
transmissions from the ground (Lawrence 2004), these observational
studies can be carried out at many wavelengths, with particular benefits in the
infrared. As the thermal background is much reduced (e.g. by a factor of 100
at 3 microns), the observational window in Antarctica is wider as compared to
tempered sites. In addition to the natural low thermal-background, recent developments
in Fiber-Bragg-Grating technologies (Bland-Hawthorn et al. 2009), allow
a sophisticated filtering of individual OH-emission lines and thus a further suppression
of the infrared sky background.

In a collaboration between the Astrophysical Institute Potsdam (AIP) and the
Physical Chemistry Group of the University of Potsdam (UPPC), interdisciplinary
research in fiber-spectroscopy and sensing (\cite{Roth2008}) as well as
in astrophotonics and astrochemistry is being combined. The aims are twofold:
to develop new astrophysical instrumentation based on photonic technologies and
to study properties and reactions of relevant molecules and ions in the laboratory.
Discussed below are the potential links of these R\&D efforts with respect to an
Antarctic observatory.

\section{Astrochemistry} 
In astrochemistry, the chemical reactions of atoms, ions, molecules and dust particles,
that are present in the interstellar medium (ISM) are being studied. In
particular, in areas with low temperatures ($10~K < T < 100~K$), a relatively high
density ($>100~cm^{-3}$) of these particles exist, which leads to rich and complex
chemical reactions. More than 100 different molecules (and their respective ions)
are observable in these areas of the ISM. The study of the formation of complex
molecules or the role of dust particles as interstellar reactors are essential topics
to understand the chemical evolution, the formation of stars and planets and the
creation of the chemical building blocks of life.

An Antarctic observatory offers the option to spectroscopically detect molecules
in the ISM from the UV to THz spectral range. This topic is related to a science
case, that was identified by the ARENA submm-working group: ``Measuring the
physical and chemical properties of the interstellar medium in our Galaxy, in the
Magellanic clouds and nearby galaxies.'' 

Of particular interest are observations of Diffuse Interstellar Bands (DIBs).
These DIBs are absorption features, produced as starlight passes through interstellar
matter. Despite the fact, that the first detections of DIBs go back to the
1920s, and that about 700 features are known so far, the identification
of their nature is one of the longest-lasting problems in astronomical
spectroscopy.

In the current understanding, DIBs are probably caused by carbon chains
(\cite{Maier2004}) or Polycyclic Aromatic Hydrocarbons (PAHs), which are complex
organic molecules that are made up from 6 to 60 carbon-atoms. PAH are
present in the ISM and can satisfy the abundance requirements for the DIBs
(\cite{S-B2008}). PAH cations are known to have rich optical and IR spectra
and according to laboratory experiments at least some overlap with observed
DIBs. \cite{Fulara1993} present laboratory evidence that highly unsaturated
hydrocarbons with carbon numbers 6-12 may be the carriers of some of the DIBs
in the range 480-1000~nm. While PAH may be good candidates to explain (some)
DIB features, it is unclear, if PAHs form in stars or by ion-molecule reactions
within the ISM. Laser-based ion mobility spectrometry is a suitable experimental tool for the investigation of the formation of large anionic PAH clusters (\cite{Beitz2006}) and PAH-mediated ion-molecule reactions (\cite{Loeh2006}), that are central issues in astrochemistry. 

The detection of diffuse interstellar bands (DIBs) at 5780 and 5797~\AA~in the
Small Magellanic Cloud and the variation of the 6284~\AA~DIB toward several targets
in the Large Magellanic Cloud (\cite{Ehrenfreund2002}) is of particular interest,
given the good visibility of these targets from Antarctica.

Further spectroscopic observations and monitoring of possible variations, in
particular at unexplored areas, are needed, ``accompanied by progress in the understanding
of the physical and chemical properties of molecules and solids and
of the physical processes and chemical reaction rates in the interstellar medium, 
by means of theoretical calculations and laboratory experiments.'' (see Astronet
Science Vision 4.3.3.)

\section{Astrophotonics} 
Astrophotonics is a relatively new research field with the aim to meet the increasing
requirements towards astronomical instrumentation. Astrophotonics investigates
new technologies (for a review see \cite{BH-K2009}), to improve and
miniaturise instruments by replacing classical optical components with photonic
devices. 

Major developments during the last years are based on the manipulation and
application of light-guides and optical fibres. Amongst others, these R\&D efforts
include photonic crystal fibres and multi-mode to single-mode couplers in ``photonic
lanterns'' (\cite{Noordegraaf2009}). Using lasers, it is possible to inscribe
refractive index changes into optical fibres (\cite{T-K-AS2009}) 
and thus to create periodic structures along the fiber axis. These Fiber-Bragg
Gratings (FBGs) act as highly complex and sophisticated filters. A demonstrated
application in astronomy is the filtering of unwanted atmospheric OH-emission
lines with the effect to reduce the IR-sky background (\cite{BH-2009}).
The use of the FBG-technology to eliminate individual emission lines, combined
with the natural low thermal background, makes an Antarctic observatory equipped
with instrumentation featuring OH-surpression fibres a unique facility to spectroscopically
observe the NIR-universe.

In another development, the properties of lasers and photonic crystal fibres
are combined to create super-continuum white light sources and laser-frequency
combs (\cite{Mandon2007}). The latter allows precise and long-term stable calibration
devices for high-resolution spectroscopy (\cite{Murphy2007}). Any dedicated
spectrograph at an Antarctic observatory, that either targets the detection
of extra-solar earths or tries to measure the expansion of the universe directly,
would require such advanced calibration technology.

Planar waveguides offer the potential to develop Integrated Photonic Spectrographs
(\cite{BH-H2006}). These miniaturised spectrographs ``on a
chip'' would be much easier to transport, to install, to thermalize and operate at
Antarctic conditions than classical bulky optics. Given their small sizes, they also
could be replicated in numbers, creating a high-multiplex factor without significantly
adding to the volume or weight budgets or the transport costs.

In addition, the above mentioned photonic technologies can be incooperated 
into fiber-based multi-object or integral-field spectrographs and thus combine
the general advantages of imaging spectroscopy (\cite{Kelz2007}) with OH-surpression
(\cite{Ellis2009}), and possibly, frequency comb calibration units to create ultra-stable,
spectrographs with high-multiplex for Antarctica. While offering almost spacelike
conditions in certain wavelength regimes, that are important for the study
of astro-chemical processes, these facilities also benefit from ground-based advantages,
such as lower costs, shorter implementation schedules, accessibility for
maintenance, repairs and upgrades to ensure state-of-the-art instrumentation.

\section{Summary}
An Antarctic observatory can serve as complementary facility to much larger
projects, such as SOFIA, ALMA, Herschel, SKA to study molecules and chemical
processes in the ISM across a wide wavelength range. On the other hand, offers the
development of astrophotonical technologies particular benefits for any Antarctic
observatory, such as miniaturisation of instrumentation or OH-surpression filters.

The interdisciplinary approach at Potsdam targets research 
in both areas of astrophotonics and astrochemistry. It is, in accordance with considerations
of the Astronet Science Vision and the Infrastructure Roadmap, aimed
to combine observations from the UV to the THz spectral range with experimental
techniques and theory to investigate ISM-relevant molecules and chemical reactions. \\

\noindent
AK gratefully acknowledges support from ARENA to attend this conference.


\end{document}